\documentclass[12pt]{article}
\usepackage{epsfig}
\title{Superintegrable models of Winternitz type.}
\author{C.Gonera\thanks{Supported by KBN grant 2 P03B 134 16}, P.Kosi\'nski$^*$, P.Ma\'slanka$^*$ \\
Dept. of Theor. Phys. II \\
Univ. of Lodz, Pomorska 149/153\\
90-236 Lodz, Poland.}
\begin{document}
\maketitle
\begin{abstract}
A general procedure is outlined which allows to construct superintegrable models
of Winternitz type. Some examples are presented.
\end{abstract}
 Due to their exceptional properties the superintegrable systems, both classical and quantum, 
are subject of constant interest. The number of examples is here, however, slightly limited.
The aim of the present paper is to outline some general procedure leading to superintegrable
models generalizing Winternitz system \cite{b1} $\frac{.}{.}$\ \cite{b4}.\\
Let us consider an integrable classical system of $N$\ degrees of freedom coupled by confining 
forces. According to the general theory \cite{b5} one can then define action-angle variables $J_{k},
\varphi_{k},\;\;k=1,...,N;$\ the submanifolds $J_{k}=const$\ are (sum of) invariant Liouville-Arnold 
tori $T^{N},$\ parametrized by the angels $\varphi_{k}$\ provided the identification 
$T^{N}\sim (S^{1})^{N}$\ has been made.\\
Generically, when restricted to invariant torus, the dynamics appears to be ergodic. If this is
not the case the system posses an additional integral of motion which is independent of $J_{k}$'s.
Such a system is called superintegrable. The maximal number of these additional integrals is 
$N-1$\ and the corresponding dynamics is called maximally superintegrable.\\
It is easy to see that all trajectories of maximally superintegrable system, being compact
common intersections of $2N-1$\ hypersurfaces, are closed. This is possible if and only if all 
ratios $\omega_{k}(J)/\omega_{j}(J)$\ of the frequencies $\omega_{k}(J)\equiv \frac{\partial
H}{\partial J_{k}}$\ are rational numbers.This implies\\
\begin{eqnarray}
\omega_{k}(J)=m_{k}\omega (J) \label{w1}
\end{eqnarray}
where $m_{k}$\ are integers and $\omega (J)$\ is a fixed function of action variables.\\
The general form of the hamiltonian leading to (\ref{w1}) is 
\begin{eqnarray}
H(J)=H(\sum_{k=1}^{N}m_{k}J_{k}) \label{w2}
\end{eqnarray}
The additionals integrals of motion can be now easily constructed \cite{b6}. To this end let $m$\ 
be the least common multiple of $m_{k},\;k=1,...,N$\ and
let $l_{k} \equiv m/m_{k}.$\ Then $\sin(l_{k}\varphi_{k}-l_{1}\varphi_{1}),
$\ (or cosines),$k=2,...,N,$\ are well-defined isolating independent integrals of motion;indeed,their time-independence is a direct consequence of eq.(\ref{w1})
while their single-valudness follows from the invariance under substitutions
\begin{displaymath}
\varphi_{i}\longrightarrow\varphi_{i}+2\Pi n_{i}, n_{i} \in Z
\end{displaymath}
 Moreover,they
are functionally independent exept nowhere dense set of points where some
of the action variables vanish.It is easy to see that ,together with the
action variables,these additional integrals define trajectories so any other
time-independent integral is expressible in terms of them.This can be checked
explicitly in the case of most prominent examples of superintegrable systems
like Kepler problem, where the additional integrals are provided by Runge-Lenz
vector, or Winternitz system.\\
As in the case of Winternitz system let us now start with completely separated hamiltonian
\begin{eqnarray}
H=\sum_{k=1}^{N}(\frac{P_{k}^{2}}{2\mu_{k}}+U_{k}(x_{k}))\equiv \sum_{k=1}^{N}H_{k} \label{w3}
\end{eqnarray}
which is immediately known to be integrable.
 However, even being so simple, $H$\ is generically not superintegrable. So one can pose the question which potentials $U_{k}(x)$\ lead to
 superintegrable dynamics, in particular-maximally superintegrable.
 Taking into account that $H$\ is completely separated and using eq.(\ref{w2}) we conclude that $H$\ should be of the form
\begin{eqnarray}
H=\alpha \cdot \sum_{k=1}^{N}m_{k}J_{k} \label{w4}
\end{eqnarray}
with some real constant $\alpha.$\ This implies that all periods
\begin{eqnarray}
T_{k}\equiv 2\pi \frac{dJ_{k}}{dH_{k}}=\frac{2\pi}{\alpha m_{k}} \label{w5}
\end{eqnarray}
are constant, i.e. energy-independent. This result is rather obvious: consider $\omega_{k}(E_{k})/\omega_{1}(E_{1})$\ as a function of $E_{k}$\ (or $E_{1})$; it is continous and attains only rational values so it must be a constant.\\
We conclude that $H,$\ given by eq.(\ref{w3}), is maximally superintegrable if
 and only if all $U_{k}(x)$\ are such that the corresponding periods
 of one-dimensional motions are energy-independent and their ratios are
 rational numbers. The solution to this  problem is, however, well known
 \cite{b6}. Assume that $U(x)$\ is such that $(i)\; 0$\ is the absolute
 minimum of $U(x),(ii)\;U(x)=E$\ has exactly two solutions $x_{1,2}(E)$\
 for any $E>0.$\ Then, given the period $T(E)$\ as a function of energy, one can find all $U(x)$, which produce $T(E),$\ from the equation
\begin{eqnarray}
x_{2}(E)-x_{1}(E)=\frac{1}{\pi \sqrt{2\mu}}\int_{0}^{E}\frac{T(\varepsilon)d\varepsilon}{\sqrt{E-\varepsilon}} \label{w6}
\end{eqnarray}
In our case $T_{k}(\varepsilon)=2\pi /\alpha m_{k}$\ so $U_{k}(x)$\ is given by 
\begin{eqnarray}
x_{2}(E)-x_{1}(E)=\frac{2}{\alpha m_{k}\sqrt{2\mu_{k}}}\int_{0}^{E}\frac{d\varepsilon}{\sqrt{E-\varepsilon}}= \frac{4}{ \alpha   m_{k} \sqrt{2\mu_{k}}}\sqrt{E} \label{w7}
\end{eqnarray}
Eq.(\ref{w7}) gives the general solution to the problem which hamiltonians $H$, eq.(\ref{w3}), are maximally superintegrable.\\
To find some particular classes of potentials $U_{k}$\ let us note that,
due to the fact that there are exactly two solutions to $U_{k}(x)=E$\ one can
write
\begin{eqnarray}
x_{2}(E)=\varphi_{k}(x_{1}(E)) \label{w8}
\end{eqnarray}
with some, yet unspecified,  function $\varphi_{k}(x).$\ \\
Then  $U_{k}(x)$\ takes the form 
\begin{eqnarray}
U_{k}(x)=\beta^{2}_{k}(\varphi_{k}(x)-x)^{2},\;\;\beta_{k}^{2}\equiv \frac{\alpha^{2}m_{k}^{2}\mu_{k}^{2}}{8}\label{w9}
\end{eqnarray}
The condition $U_{k}(x)=U_{k}(\varphi_{k}(x))$\ will be obeyed if $\varphi_{k}(\varphi_{k}(x))=x.$\ The choice $\varphi_{k}(x)=-x$, resp. $\varphi_{k}(x)=\gamma_{k}/x$, correponds to harmonic oscillators, resp. Winternitz system.\\
It is not difficult to find other examples. Let $\eta_{k}>0$\ be arbitrary real positives (of dimension of lenght) and let $\varphi_{k}(x)$\
\begin{eqnarray}
\varphi_{k}(x)=\frac{\eta_{k}x}{\sqrt{x^{2}-\eta^{2}_{k}}}\label{w10}
\end{eqnarray}
Then $\varphi_{k}\circ \varphi_{k}=id$\ and $U_{k}(x)$\ defined by (\ref{w9}) obeys $(i),(ii).$\ Therefore, the hamiltonian 
\begin{eqnarray}
H=\sum_{k=1}^{N}(\frac{p_{k}^{2}}{2\mu_{k}}+\beta_{k}^{2}x_{k}^{2}(1-\frac{1}{\sqrt{(\frac{x_{k}}{\eta_{k}}^{2})^{2}-1}})^{2})\label{w11}
\end{eqnarray} 
with $\eta_{k}<x_{k}< \infty$, is maximally superintegrable.
In principle, one can construct the additional integrals of motion
according to the recipe formulated above. However, they are not expressible
in terms of simple (elementary, elliptic...) functions. This model can be
 easily generalized. Take $n_{k}$\ to be positive integer and write
\begin{eqnarray}
\varphi_{k}(x)\equiv \eta_{k}x(x^{2n_{k}}-\eta_{k}^{2n_{k}})^{-\frac{1}{2n_{k}}},\;\;\eta_{k}<x< \infty \label{w12}
\end{eqnarray}
Again one checks easily that $\varphi_{k}\circ \varphi_{k}=id$\ and
$U_{k}(x)$\ obeys $(i)$\ and $(ii).$\ \\
Although the explicit expression for additional integrals of motion are
not accessible it is not difficult to generalize the Evans result
\cite{b4} concerning the dynamical Poisson algebra for Winternitz system.
It appears that for the systems defined above it is again $sp(2N,R)
$\ \cite{b7}.\\
We have assumed above that there are exactly two solutions to the equation
$U_{k}(x)=E$ because for this case the problem of finding $U_{k}(x)$ in
terms of $T_{k}(E)$ is tractable in a simple way.However,this assumption
seems to be also important on its own because we need $T_{k}$ to be energy-
independent.For the energies E close to the local minimum the frequency
squared of the motion equals second derivative of the potential at the minimum.
On the other hand,if $U(x)=E$ has more solutions there must be also local
maximum of $U(x)$ and for the energy equal to the value of $U(x)$ at such
a maximum the period becames infinite.In other words,the period can not
be energy-independent.\\ 
The algorithm outlined above allows to produce many superintegrable systems of arbitrary number 
of degrees of freedom. It allows also in principle to construct the additional integrals of motion; one has to find the angle variables and to construct the appropriate single-valued functions of them.\\
It is also worth to note that eq.(\ref{w6}) allows to construct the
models which are superintegrable in some region of phase space only.
To this end we select for any $k$\ the energy intervals
$E_{k},E_{k}+\Delta_{k}$\ and assume that
$T_{k}(E_{k})\equiv T_{k}$\ are constant over intervals and $T_{k}/T_{l}$\ are rational. Eq.(\ref{w6}) allow us then to find the relevant potentials.
The resulting hamiltonian is superintegrable in the above specified region of phase space.

\end{document}